\documentclass[twocolumn,showpacs,preprintnumbers,amsmath,amssymb]{revtex4}

\usepackage{graphicx}
\usepackage{dcolumn}
\usepackage{bm}

\begin{document}

\preprint{APS/123-QED}

\title{Synchronization of time delay systems using variable delay with reset for enhanced security in communication}

\author{G. Ambika}
\email{g.ambika@iiserpune.ac.in}
\affiliation{ Indian Institute of Science Education and Research, Pune 411 021, India}
\author{R.E. Amritkar}
\email{amritkar@prl.res.in}
\affiliation{Physical Research Laboratory, Ahmedabad 380 009, India}

\date{\today}

\begin{abstract}
 We have introduced a mechanism for synchronizing chaotic systems by one way coupling with a variable delay that is reset at finite intervals. Here we extend this method to time delay systems and suggest a new cryptosystem based on this. We present the stability analysis as applied to time delay systems and supplement this by numerical simulations in a standard time delay system like Mackey Glass system.  We extend the theory to multi- delay systems and propose a bi channel scheme for the implementation of the scheme for communication with enhanced security. We show that since the synchronizing channel carries information from transmitter only at intervals of reset time, it is not susceptible to reconstruction. The message channel being separate can be made complex by linear combination of transmitter variable at different delay times using mutiple delay systems.  This method has the additional advantage that it can be adjusted to be delay or anticipatory in synchronization and these provide two additional basic keys that are independent of system delay.
\end{abstract}

\maketitle

\section{Introduction}
In recent years, a number of cryptosystems based on synchronized chaotic systems  have been introduced \cite{Yang04,Koc95}. Many of these systems generally make use of the broad band  spectrum of the chaotic output from the transmitter as a carrier for secure masking of the signal and its synchronization with the receiver for easy recovery of the signal.  In addition to chaotic masking, several other schemes involving chaotic modulation and chaotic switching etc.  are also used for a practical realization of the chaos based cryptosystem \cite{Chi05,Kle05,He98}.  It has been shown that message transmission using synchronization  can be implemented in practice using semi conductor lasers and electronic circuits of chaotic systems \cite{Cuo93,Wag08,Yang97}. The most frequently studied schemes involve complete synchronization of one way coupled identical systems. However generalized synchronization of non identical systems and phase synchronization have been adapted to secure communication using Chua's circuits \cite{Mur98,Chen03}.

Of late, it has been shown that low dimensional chaotic systems can be easily unmasked by reconstructing the dynamics from time series available from the channel by an intruder. Since then, the use of hyper chaotic systems derived from time delay systems have been proposed for improved security \cite{Pyr98}. Their methods of implementation using time delayed optical systems and circuits and generation of secret keys have also been studied in detail\cite{Van99,Li04,Cruz04,Oht02,Ban09,Kim04,Lar04,Lar05,Arg05,Lav09}. However, later studies have indicated ways of reconstruction of such systems from time series and extraction of messages masked by hyper chaotic carriers \cite{Bez,Pon02,Zhou99,Uda03,Uda05,Yan04}. In general, such cracking schemes, are aimed at identifying the time delay parameter from the cipher text in the channel. For enhanced security, several variations like multiple delays, variation and modulation of delays, harmonics  and multi channels  have been reported \cite{Lia03,Liu00, Sha09, Ron07}.  

We have recently introduced a mechanism for synchronization of chaotic systems by coupling involving a variable delay that is reset at finite intervals \cite{Amb09}. In this method, the delay varies in the same way as the system in time and hence the value of the synchronization signal remains constant for  the reset interval at the end of which it is reset to the value at that time. We have applied this to standard low dimensional systems like R\"ossler and Lorenz and carried out the stability theory and Lyapunov exponents analysis for such a coupling scheme. 

In this paper, we further develop this method to apply to hyper chaotic time delay systems like Mackey Glass and Ikeda systems with possible applications to secure communication. In this context, we note that several studies on variable and multiple  time delays in synchronized time delay systems have been reported \cite{Gju10}  with applications to communication \cite{Gho07, Sen07, Zhe08, Sou08}. But in our scheme, unlike the above methods, the variability is in the delay or anticipation time in the coupling function and not  in the feedback delay in the system dynamics. We further suggest a bi-channel scheme for its implementation where the synchronization channel carries information about the transmitting dynamics only at intervals of reset time and hence is not susceptible to reconstruction by an intruder. The message channel being separate can be made complex enough. At large reset times, the reported methods of reconstructing the dynamics by first computing the delay time of the system dynamics is not possible, since the information of the system variable is available only at intervals of the reset time.  By a suitable choice of the carrier as a combination with multiple delays from the transmitter for effective  masking, we show how the message can be retrieved truthfully at the receiver once it synchronizes with the transmitter. 

The method reported here has the additional advantage that it can be easily adjusted to be delay, anticipatory or isochronous in synchronization. Anticipation in the context of communication  is not essential even though it will give more time for decoding at the receiver. However, the delay and anticipatory times serve as additional keys. In this paper we give more emphasis on small anticipatory times so that the reset time can be maximized with large enough system delay. This makes the method highly cost effective  with more complex  output  for enhanced  security. Such optimization is possible due to the flexibility among the four different time scales in our method viz. delay in the transmitting system , anticipation in the receiving system, system delay and reset times.
We present the stability analysis as applied to time delay systems and supplement this numerically by computing the stability regions for synchronization in the parameter space for a standard time delay system normally used for communication viz. Mackey Glass (MG) system.  We also present the study of two MG systems with multiple delays using the present scheme. The applicability for communication is illustrated by a bi-channel scheme for an analogue signal. We establish the enhanced security of this method by explicitly applying the existing methods of cracking applicable to such time delay systems.
\section{Synchronization of time delay systems with variable delay and reset}

We consider systems whose intrinsic dynamics follows a time delay equation
\begin{equation}
\dot{x} = -\kappa x + f(x_{\tau_s})
\label{eq:1}
\end{equation}
where $x_{\tau_s} = x(t-\tau_s)$; $\tau_s$ being the time delay in the system. 
  
Two such systems are coupled in the drive response mode  with a linear difference coupling where the drive variable is delayed by $\tau_1$ and the response by $\tau_2$. These delays vary in time with the same time scale as the system within an interval called the reset time $\tau_r$. At the end of this interval, the delay is reset to the initial value and the dynamical variable set to its value at that time. This makes the coupling function constant over $\tau_r$ and hence the information from the transmitter for the synchronizing signal is required only at intervals of $\tau_r$. The dynamics of the two coupled systems thus evolves under four different time scales in addition to the system dynamics,  viz  the delay time $\tau_1$, the anticipatory time $\tau_2$, the reset time $\tau_r$, the system delay time $\tau_s$.  We prescribe the following dynamics for the coupled systems under this scheme. 

\begin{subequations}
\label{system_def}
\begin{eqnarray}
    \dot{x} & = & -\kappa x + f(x_{\tau_s}) 
\label{system_def_x}\\ 
    \dot{y} & = & -\kappa y + f(y_{\tau_s})\nonumber \\
&& + \epsilon \sum_{m=0}^{\infty} (x_{t_1}-y_{t_2}) \chi_{(m\tau_r,(m+1)\tau_r)}
\label{system_def_y}
\end{eqnarray}
\end{subequations}
where $x_{t_1} = x(t-t_1)$, $y_{t_2} = y(t-t_2)$, $\tau_r$ is the resetting time and $\chi_{(t',t'')}$ is an 
indicator function such that
$\chi_{(t',t'')} = 1 \; \textrm{for} \; t' \leq t \leq t''$ and
zero otherwise. Both the delays $t_1$ and $t_2$ depend on time and 
we choose this dependence as
\[ t_i = \tau_i + t - m\tau, \; \; i=1,2. \]
This takes care of the variation in the delay times mentioned above.  The synchronization manifold for the coupled systems (\ref{system_def}) is defined  by 
$y(t-\tau_2) = x(t-\tau_1) \; {\textrm or} \;
y(t) = x(t-\tau_1+\tau_2)$ 

We have shown that under this type of coupling, it is possible to have isochronal, delay or anticipatory synchronization between chaotic systems \cite{Amb09}. In all these cases, the transverse system can be defined by $\Delta = y - x_{\tau_1-\tau_2}$. Its dynamics in linear approximation can be derived from Eq.~(\ref{system_def}) as

\begin{eqnarray}
 \dot{\Delta}& = & -\kappa \Delta +f ^{'}(x_{\tau_1-\tau_2 +\tau_s}) \Delta_{\tau_s}\nonumber\\
&&-\epsilon \sum_{m=0}^{\infty} \chi_{(m\tau_r,(m+1)\tau_r)} \Delta_m 
\label{linear_stability}
\end{eqnarray}
where $\Delta_{m} = \Delta(t-t_2) = \Delta(m\tau-\tau_2)$ is a constant in any reset interval. $\tau_1$ can eliminated by shifting the time scale of drive system and redefining $\tau_2$.

We approximate the equation by replacing  $f^{'}$ by an effective time average value $\nu$ 

\begin{equation}
 \dot{\Delta} = -\kappa \Delta +\nu \Delta_{\tau_s} - 
\epsilon \sum_{m=0}^{\infty}  \chi_{(m\tau_r,(m+1)\tau_r)} \Delta_m
\label{linear_problem}
\end{equation}

Assuming a solution $ \Delta = e^{\lambda t}$ for the equation, 
\begin{equation}
 \dot{\Delta} = -\kappa \Delta +\nu \Delta_{\tau_s} 
\label{solve_1}
\end{equation}
we get
\begin{equation}
 \lambda = -\kappa + \nu e^{-\lambda \tau_s} 
\label{solve_2}
\end{equation}

\begin{figure}
\includegraphics[width=0.9\columnwidth]{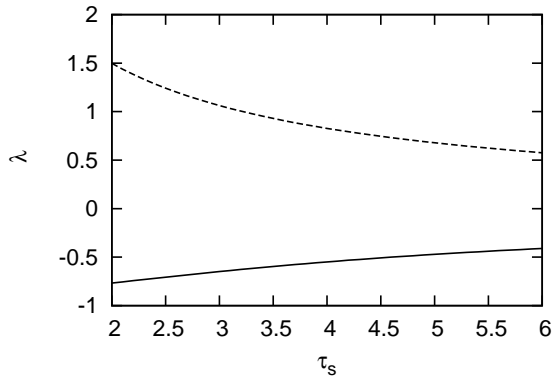}%
\caption{\label{mg-lam-ts}The variation in $\lambda$ as the system delay increases for two cases,$\kappa <\nu $(upper curve) $\kappa > \nu $(lower curve). 
}
\end{figure}

Then the solution of Eq.~(\ref{linear_problem}) in any reset interval , $m\tau \leq t < (m+1) \tau$, is
\begin{equation}
 \Delta = \alpha \Delta_m + C_m e^{\lambda t}
\label{sol-Cm}
\end{equation}
where $\alpha = \epsilon / \lambda$ is the normalized dimensionless coupling constant, and $C_m$ is an integration constant,which can be determined by matching solutions across the reset points. Then the stability analysis detailed in \cite{Amb09}, can be easily adapted to time delay systems also, the only difference being that here $\lambda$ depends on the  system time delay $\tau_s$ and has to be evaluated for each case using Eq.~(\ref{solve_2}). This transcendental equation can be solved to show that for $\nu>0 $ and $\kappa <\nu $, $\lambda$ is positive and as $\tau_s$ increases $\lambda$ decreases, while for $\kappa >\nu $, $\lambda$ is negative and as $\tau_s$ increases $\lambda$ increases. This is shown in Fig.~\ref{mg-lam-ts}.

In this paper, since we are interested in applications to communication, $\tau_2$ need not be large and we concentrate in the regime  $0 \leq \tau_2 \leq \tau_r$. Then the upper limit of stability in the parameter plane $\tau_2/\tau-\alpha$ as derived in \cite{Amb09} has a peak at $\tau_{2p}$. For $\tau_2 \leq \tau_{2p}$, the upper limit of stability is given by 
\begin{equation}
 \alpha_u = \frac{e^{\lambda \tau_r}+1}{2e^{\lambda (\tau_r- \tau_2)} - e^{\lambda \tau_r} - 1}
\label{alphau1}
\end{equation}
while for larger values of $\tau_2$ ($\tau_{2p} \leq \tau_2 \leq \tau_r$) it is given by
\begin{equation}
 \alpha_u = \frac{e^{-\lambda \tau_r}}{1-e^{-\lambda \tau_2}}
\label{alphau2}
\end{equation}

Moreover, in the present context, the maximum reset time for a given $\tau_2$ may be more relevant and from our earlier analysis, it is  

\begin{equation}
 \lambda \tau_{rmax} = -\ln (1-e^{-\lambda \tau_2}).
\label{taurmax-0} 
\end{equation}

Using Eq.~(\ref{solve_2}) and Eq.~(\ref{taurmax-0}), $\tau_{rmax} $ for given $\tau_2$ and for different $\tau_s$ can be calculated.  

We further extent this scheme to systems with multiple delays that are more realistic models in many applications. They occur in many situations, an example for such multi-delay systems being the semiconductor laser with multiple opto electronic feed backs, where additional feed back is often required to stabilize the output. Moreover such systems are of practical importance especially in the context of communication where the higher complexity can lead to better message security. The dynamics for two such systems coupled using the present scheme is  

\begin{subequations}
\label{msystem_def}
\begin{eqnarray}
    \dot{x} & = & -\kappa x + \sum_{i=0}^ N a_i f(x_{\tau_{s_i}}) 
\label{msystem_def_x}\\ 
    \dot{y} & = & -\kappa y + \sum_{i=0}^ N a_i f(y_{\tau_{s_i}})\nonumber\\
&& + \epsilon \sum_{m=0}^{\infty} (x_{t_1}-y_{t_2}) \chi_{(m\tau_r,(m+1)\tau_r)}
 \label{msystem_def_y}
\end{eqnarray}
\end{subequations}

A similar analysis will furnish the stability criteria with $\lambda$ replaced as 

\begin{equation}
 \lambda = -\kappa + \sum_{i=0}^ N a_i \nu_i e^{\lambda \tau_{s_i}} 
\label{msolve_0}
\end{equation}

The above analysis is supported by numerical simulations using a standard time delay system, Mackey Glass system, in the next section.

\section{Mackey Glass systems with variable delay in coupling}

 The Mackey Glass system, first introduced as a model for blood generation in patients with leukemia, is well studied as a model exhibiting hyper chaos \cite{Pyr98pre}. The synchronization in two such coupled systems has been reported for various types of synchronization \cite{Chen07}. It has been applied extensively in the context of communications, especially with  its circuit equivalents \cite{Sano07}. The synchronization regimes of two such systems with multiple delays have been reported recently \cite{Tha07}.  So also coupling with multiple delays \cite{Hoa05,Sha06} has been proposed for more secure communications. In this section, we apply the scheme of variable delay described in the last section  to two coupled Mackey Glass systems. The dynamics of a single system is

\begin{equation}
 \dot{x} = -a x + \frac{b x (t-\tau_s)}{1+x(t-\tau_s)^c}
\label{MG_1}
\end{equation}

As reported in \cite{Pyr98pre}, for a=1, b=2 and c=10, as the delay time $\tau_s$ is varied, the system has a fixed point attractor for $\tau_s < 0.471$, a limit cycle for $0.417<\tau_s<1.33 $, a period doubling sequence in the range $\tau_s = [1.33, 1.68]$ and chaos for $\tau_s >1.68 $. Thereafter the number of positive Lyapunov exponents increases linearly with $\tau_s$, and the system shows hyper chaotic behavior. In our simulations here, we choose a similar set of parameter values with $\tau_s$ between 2.5 and 5.0 corresponding to the hyper chaotic region and couple two such systems using our scheme given in Eq.~\ref{system_def} as 

\begin{subequations}
\label{MG2_def}
\begin{eqnarray}
    \dot{x} & = & - x + \frac{2 x (t-\tau_s)}{1+x(t-\tau_s)^{10}}
\label{MG2_def_x}\\ 
    \dot{y} & = & - y + \frac{2 y (t-\tau_s)}{1+y(t-\tau_s)^{10}}\nonumber\\
&& + \epsilon \sum_{m=0}^{\infty} (x_{t_1}-y_{t_2}) \chi_{(m\tau_r,(m+1)\tau_r)}
 \label{MG2_def_y}
\end{eqnarray}
\end{subequations}

\begin{figure}
\includegraphics[width=0.9\columnwidth]{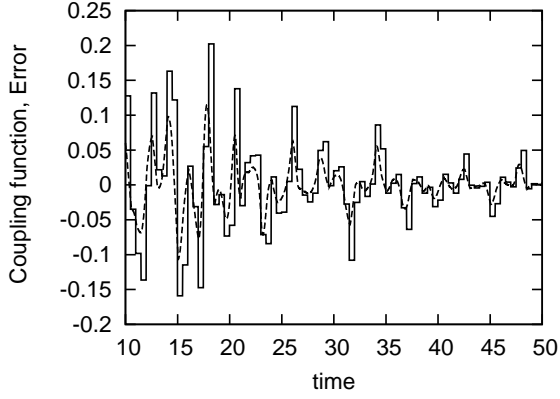}%
\caption{\label{mg-CF-er}The nature of variation of the coupling function having variable delay with reset(full line) and the synchronization error(dashed curve) for two coupled Mackey Glass systems. Here $\tau_s $=2.5, $\tau_r$= 0.5, $\tau_1$=.02, $\tau_2$= 0.1 and $\epsilon$ = 2.0 
}
\end{figure}

For a typical choice of parameters, $\tau_s$ = 2.5, $\tau_r$= 0.5, $\tau_1$=.02, $\tau_2$= 0.1, and coupling strength, $\epsilon$ = 2.0, we present the variation in the coupling function according to our scheme and the corresponding error function in Fig.~\ref{mg-CF-er}. It is clear that the coupling function is not changing within each reset interval and the synchronization error approaches zero.

\begin{figure}
\includegraphics[width=0.9\columnwidth]{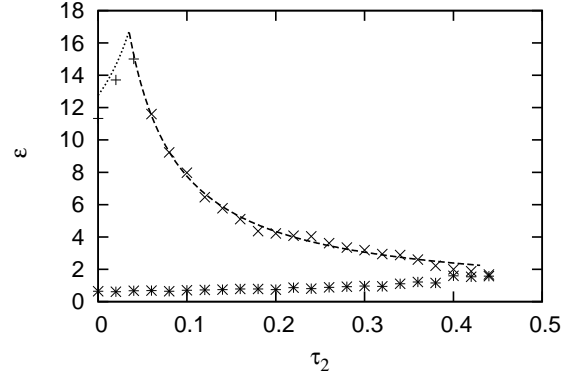}%
\caption{\label{mg-C-t2}The limits of stability of the synchronized state for two coupled Mackey Glass systems in the parameter plane $\tau_2-\epsilon$. The dotted and dashed curves for the upper boundary correspond to the curves obtained by fitting Eq.~\ref{alphau1} and Eq.~\ref {alphau2} obtained from the stability analysis, while the points correspond to the values obtained by direct numerical simulations using the correlation index. Here $\tau_s $=2.5, $\tau_r$= 0.2 and $\tau_1$=.02. 
}
\end{figure}

By numerically integrating the coupled systems in Eq.~\ref{MG2_def}, we calculate the correlation coefficient  $C=<y_1(t)x_1(t+\tau_2)>/\sqrt{<x_1^2(t)><y_1^2(t)>}$ between $x_1(t)$ and $y_1(t)$ shifted by  the effective $\tau_2 = | \tau_2-\tau_1|$ (hereafter referred to as $\tau_2$ itself).The region of stability of the synchronized state  is isolated as the region where $C= 0.999$ and boundaries of stability fixed when $C$ goes below this value.The stability region thus obtained in the parameter plane $\tau_2-\epsilon$ is plotted in Fig.~\ref{mg-C-t2}. The region below the upper curve and above the lower curve corresponds to total synchronization with correlation function $ ~ 1.0$ . The dotted and dashed curves for the upper boundary correspond to the curves obtained by fitting Eq.~\ref{alphau1} and Eq.~\ref {alphau2} obtained from the stability analysis, while the points correspond to the values obtained by direct numerical simulations using the correlation index. The agreement is found to be very good. 

\begin{figure}
\includegraphics[width=0.9\columnwidth]{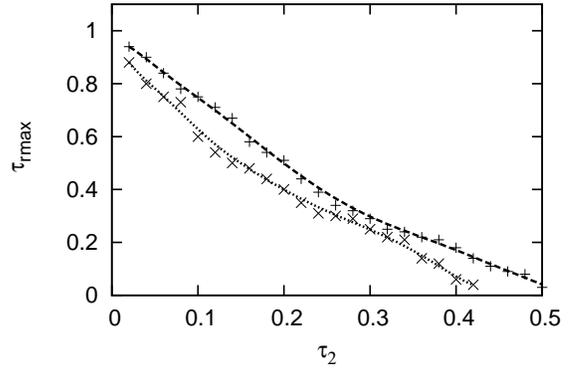}%
\caption{\label{mg-trmax-t2}The maximum reset time for stability for a given $\tau_s$ for two coupled Mackey Glass systems as a function of $\tau_2$ for two values of $\tau_s$, 2.5 and 5.0. 
}
\end{figure} 

The maximum reset time for a given $\tau_s$ obtained by numerical calculations is given in Fig.~\ref{mg-trmax-t2} for two different values of $\tau_s$, 2.5 and 5.0. It is interesting that as $\tau_s$ increases, maximum reset time also increases for a given $\tau_2$.  This is a desirable feature in applications like communication. 

We extend our numerical computations to MG systems with multiple delays. The dynamics of two such systems with four delays is given by 
\begin{subequations}
\label{mMG2_def}
\begin{eqnarray}
    \dot{x} & = & - x + \sum_{i=1}^ 4 \frac{b_i x (t-\tau_{s_i})}{1+x(t-\tau_{s_i})^{10}}
\label{mMG2_def_x}\\ 
    \dot{y} & = & - y + \sum_{i=1}^ 4 \frac{b_i y (t-\tau_{s_i})}{1+y(t-\tau_{s_i})^{10}}\nonumber\\
&& + \epsilon \sum_{m=0}^{\infty} (x_{t_1}-y_{t_2}) \chi_{(m\tau_r,(m+1)\tau_r)}
 \label{mMG2_def_y}
\end{eqnarray}
\end{subequations}

\begin{figure}
\includegraphics[width=0.9\columnwidth]{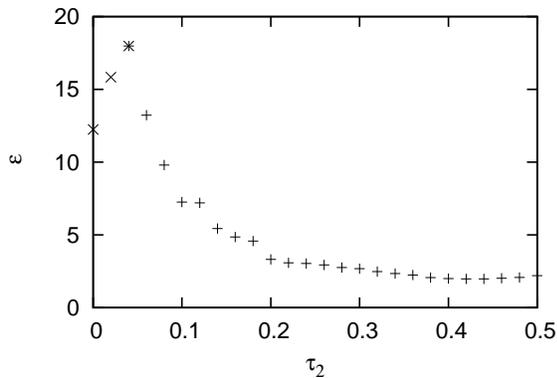}%
\caption{\label{mg-C-t2-4d}The stability boundary for two coupled Mackey Glass systems with 4 delays in the parameter plane $\tau_2-\epsilon$ obtained by direct numerical simulations using the correlation index. Here $\tau_r$= 0.5 and $\tau_1$=.02.
}
\end{figure}

The stability region in parameter plane, $\tau_2-\epsilon$ obtained by direct numerical analysis of Eq.~\ref {mMG2_def}, using the same correlation index as above is given in Fig.~\ref{mg-C-t2-4d}. Here $(b_i,\tau_{s_i})$ values used are (2.0,2.5),(0.2,2.2),(0.05,2.3)and (0.1,2.4). We note that the region of stability in this case is more compared to single delay which may be useful in many applications.

\section{ Application to secure communication}
In this section we propose a scheme for secure communication using hyper chaotic systems using the method of synchronization discussed in this paper. We also show how the method can lead to enhanced security.  We follow the chaotic masking  method as applied to a bi-channel scheme. The transmitter and receiver dynamics in our scheme are given by Eq.~\ref{mMG2_def}. Here we take $(b_i, \tau_{s_i})$ as (2.0,5.0), (0.2,4.0),(0.05,3.0),(0.1,2.0).
The synchronizing signal S(t) is the intermittent values of x(t- $\tau_1$) sent at intervals of the reset time $\tau_r$.

The message to be transmitted is assumed to be a sine signal given by

\begin{equation}
 m(t) = 0.01 sin (\omega t)
\label{sig_1}
\end{equation}

This message is encrypted using method given in \cite{Tha07} for bi-channel scheme. The  carrier is generated by combining the output of the transmitter at eleven different delay times using a nonlinear function as given below. 

\begin{equation}
 CW(t) = \sum_{i=0}^ {10} \frac{ a_i x (t-\tau_{s_i})}{1+x(t-\tau_{s_i})^ {10}}
\label{car_1}
\end{equation}
Here the values of $(a_i, \tau_{s_i})$ used are (4.0,2.0,), (5.0,1.9), (6.0,1.8), (7.0,1.7), (8.0,1.6), (9.0,1.5), (10.0,1.4), (11.0,1.3),(12.0,1.2), (13.0,1.1), (14.0,1.0).
This carrier is used to mask the message and the cipher text C(t) thus obtained is 
\begin{equation}
 C(t) = m(t) + \sum_{i=0}^ {10} \frac{ a_i x (t-\tau_{s_i})}{1+x(t-\tau_{s_i})^{10}}
\label{cipher_1}
\end{equation}

\begin{figure}
\includegraphics[width=0.9\columnwidth]{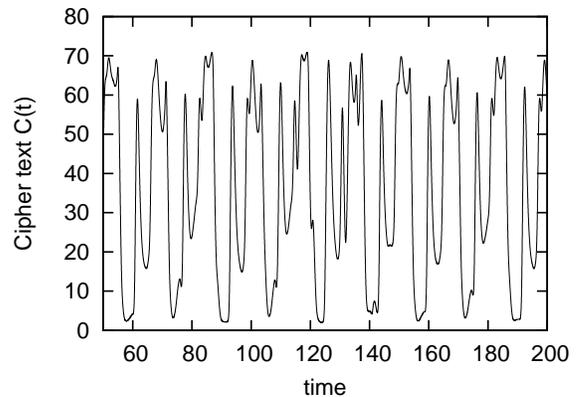}%
\caption{\label{mg-cipher-4d}The cipher text CT(t) given in Eq.\ref{cipher_1} generated by masking the signal m(t) using carrier wave given in Eq.~\ref{car_1}. 
}

\end{figure}

\begin{figure}
\includegraphics[width=0.9\columnwidth]{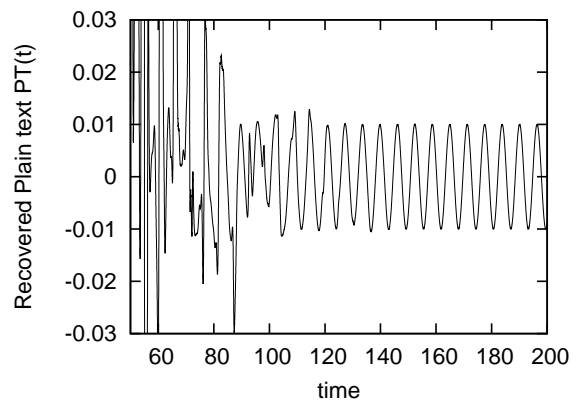}%
\caption{\label{mg-rec-4d}The plain text PT(t) or recovered signal using Eq.\ref{plain_1}.
}
\end{figure}

\begin{figure}
\includegraphics[width=0.9\columnwidth]{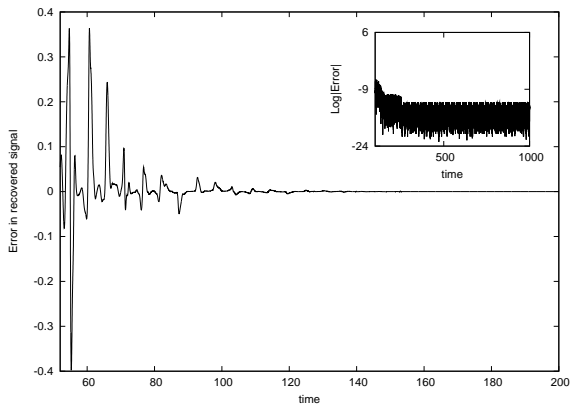}%
\caption{\label{mg-er-4d}The error in the recovered plain text as compared to the message m(t). The inset shows the $Log|error|$ for longer times.
}
\end{figure}

This  cipher text is communicated to the receiver along one channel, while the synchronizing signal S(t) is transmitted through another channel. Once the systems  at both ends are synchronized, the signal is recovered at the receiving end as plain text. 
\begin{equation}
 P(t) = C(t) - \sum_{i=0}^ {10} \frac{ a_i y(t-\tau_{s_i})}{1+y(t-\tau_{s_i})^{10}}
\label{plain_1}
\end{equation} 
We display the cipher text C(t), the recovered plain text P(t) and the error in the recovered plain text in Fig.~\ref{mg-cipher-4d}, Fig.~\ref{mg-rec-4d} and Fig.~\ref{mg-er-4d} computed using the method. It is seen that after the transients, the error decreases to zero and the message can be recovered truthfully. 

\begin{figure}
\includegraphics[width=0.9\columnwidth]{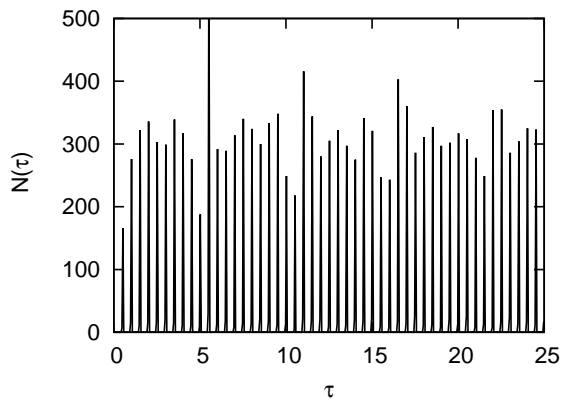}%
\caption{\label{mg-sync-his}The method reconstruction in \cite{Pon02} applied to the synchronizing signal to extract the system delay $\tau_s$. It is clear that there is no single absolute minimum. 
}
\end{figure}

To illustrate the enhanced security of the method, we note that S(t), derived using the present scheme,  has information about the dynamics of the transmitter only at intervals of rest time $\tau_r$ and as such can not be used to derive the dynamics using conventional methods of cracking reported so far. We show this explicitly by considering the method  used in \cite {Bez,Pon02}, where the time delay in the dynamics of the systems is to be extracted first to reconstruct the transmitter system. For this the number of pairs of extrema separated in time by different $\tau$ in the time series obtained from the channel is taken and the absolute minimum gives a clue to the time delay in the system $\tau_s$. Once this is obtained the  functions in the transmitter dynamics can be derived \cite{Pon02}. For applying this to our method, we reconstruct the time series from the synchronizing channel by removing the reset intervals and use this collapsed time series to take the count of extrema for different separations. This is given in Fig.~\ref{mg-sync-his}. It is clear that as the minima happen at a large number of points due to finite reset times, an estimate of the time delay $\tau_s$ cannot be obtained.    

We also use the same technique for checking possibility of reconstruction from the cipher text. Here for the same combination of eleven nonlinear functions used above for the cipher text  in Eq.~\ref{cipher_1}, we find that the information about the system delay is sufficiently hidden.(Fig.~\ref{mg-cipher-4d}). Moreover, the eleven different sets of parameters ($b_i, \tau_{s_i})$ used in constructing the cipher text gives added security, by providing extra keys. In addition to this in our method, the two time delays $\tau_1$ and $\tau_2$ serve as additional keys.
\begin{figure}
\includegraphics[width=0.9\columnwidth]{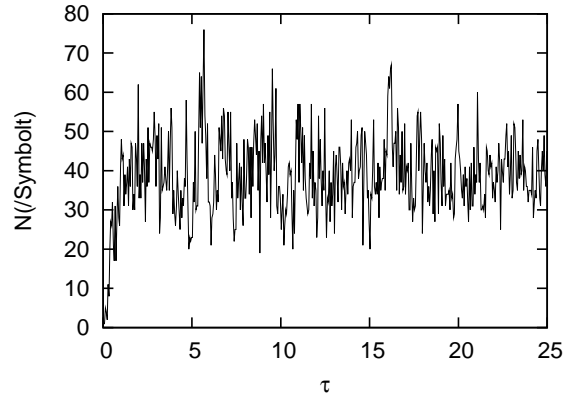}%
\caption{\label{his-4d}The method of reconstruction in \cite{Pon02} applied to the cipher text. The complexity of the cipher text gives several minima and hence it is dificult to get an indication of $\tau_s$.  
}
\end{figure}

\section{ Conclusion}

 We have introduced a coupling scheme for synchronizing two one way coupled chaotic systems where the delay in coupling varies with system dynamics within intervals of the reset time. In this paper, we have extended this scheme to include time delay systems where the synchronization can be delay or anticipatory independent of the system delay. The stability analysis is adapted for such systems and applied to a standard time delay system like Mackey Glass system. We suggest the possibility of applying this method in communication with enhanced security. For this, we propose a bi channel method for implementation of the same. Our method has the advantage that synchronizing channel carries only intermittent information from the driver. We show that this enhances the security of transmission as extracting system delay from this channel is not possible by the conventional methods reported so far. The cipher text channel can be made complex resulting in extra keys. We illustrate this by using a carrier wave constructed from a multi delay Mackey Glass system as driver with different delays. Such bi channel schemes have been reported earlier also. But our scheme promises extra security in the synchronizing channel due to intermittent transmission and the delay and anticipation in synchronization in  our method function as additional keys that are independent of system delay.


\begin{thebibliography}{1}
\bibitem{Yang04} Tao Yang, International Journal of Computational Cognition, {\bf 2}, 81 (2004)
\bibitem{Koc95} L. Kocarev and U. Parlitz, Phys. Rev. Lett. {\bf 74}, 5028 (1995)
\bibitem{Chi05} Tsun-I Chien and Teh-Lu Liao, Chaos, Solitons \& Fractals {\bf 24}, 241 (2005). 
\bibitem{Kle05} E. Klein, R. Mislovaty, I. Kanter and W. Kinzel, Phy. Rev. {\bf E 72}, 016214 (2005).

\bibitem{He98} R. He and P. G. Vaidya, Phy. Rev. {\bf E 57}, 1532  (1998).
\bibitem{Cuo93} K. M. Cuomo,  A. V. Oppenheim and S. H. Strogatz, IEEE Transactions on Circuits \& Systems-11: Analog \& Digital Signal Processing {\bf 40}, 626 (1993). 
\bibitem{Wag08} A. Wagemakers, J. M. Buld'u and M. A. F. Sanju'an, Eur.Phy. Lett. {\bf 81}, 40005 (2008) 
\bibitem{Yang97} T. Yang and L. O. Chua, IEEE Transactions on Circuits \& Systems-I: Fundamental Theory \& Applications {\bf 44}, 976 (1997)
\bibitem{Mur98} K. Murali  and M. Lakshmanan, Phys. Lett. {\bf A 241}, 303 (1998) 

\bibitem{Chen03} J.Y.Chen, K. W. Wong, L. M. Cheng and J. W. Shuai ,CHAOS {\bf 13}, 508 (2003.


\bibitem{Pyr98}  K. Pyragas, Int. J. Bif. \& Chaos {\bf 8}, 1839 (1998). 

\bibitem{Van99} G. D. Vanwiggeren and R. Roy, Int. J. Bif. \& Chaos {\bf 9}, 2129 (1999).

\bibitem{Li04} C. Li, X. Liao and  K. Wong, Physica {\bf D 194}, 187 (2004).

\bibitem{Cruz04} C. Cruz-Hern'andez, Nonlinear Dynamics and Systems Theory {\bf 4}, 1 (2004).

\bibitem{Oht02} J. Ohtsubo, IEEE Journal of Quantum Electronics {\bf 38}, 1141 (2002). 

\bibitem{Ban09} S. Banerjee, Chaos , Solitons \& Fractals {\bf 42}, 745 (2009).

\bibitem{Kim04} Chil-Min Kim , Won-Ho Kye, Sunghwan Rim and Soo-Young Lee, Phys. Lett. {\bf A 333}, 235 (2004).

\bibitem{Lar05} L. Larger, P.A. Lacourt,  S. Poinsot,  and V. Udaltsov, Laser Physics {\bf 15}, 1209 (2005).
\bibitem{Lar04} L. Larger, J. P. Goedgebuer and V. Udaltsov, C. R. Physique {\bf 5}, 669 (2004) 

\bibitem{Arg05} A. Argyris, D. Syvridis, L. Larger, V. Annovazzi-Lodi, P. Colet, I. Fischer, J. Garci­a-Ojalvo, C. R. Mirasso, L. Pesquera and K. A. Shore, Nature {\bf 438}, 343 ( 2005) 

\bibitem{Lav09} R. Lavrov, M. Peil, M. Jacquot, L. Larger, V. Udaltsov, and J. Dudley, Phy. Rev. {\bf E 80}, 026207 (2009).

\bibitem{Bez} B. P. Bezruchko, A. S. Karavaev, V. I. Ponomarenko, and M. D. Prokhorov, Phy. Rev. {\bf E 64}, 056216 (2001)

\bibitem{Pon02} V. I. Ponomarenko and M. D. Prokhorov, Phys. Rev. {\bf E 66}, 026215 (2002).

\bibitem{Zhou99} C. Zhou  and C.H. Lai, Phy. Rev. {\bf E 60}, 320 (1999).
\bibitem{Uda03} V. S. Udaltsov , Jean-Pierre Goedgebuer , L. Larger , Jean-Baptiste Cuenot ,P. Levy and W. T. Rhodes, Phys. Lett. {\bf A 308}, 54 (2003).
\bibitem{Uda05} V. S. Udaltsov, L. Larger,  J. P. Goedgebuer , A. Locquet and D. S. Citrin,  J. Opt. Technol. {\bf 72}, 373 (2005).
\bibitem{Yan04} Z. Yan, L. Yaowen, and W. Yinghai, Chinese Journal of Physics, {\bf 42}, 323 (2004).
\bibitem{Lia03} L. Wu and S. Zhu, Phys. Lett. {\bf A 308}, 157 (2003.

\bibitem{Liu00}  LiuYaowen, GeGuangming, ZhaoHong, WangYinghai, GaoLiang, Phy. Rev. {\bf E 62}, 7898 (2000).

\bibitem{Sha09} E.M. Shahverdiev and K.A. Shore, Optics Communications {\bf 282}, 3568 (2009).
\bibitem{Ron07} D. Rontani, A. Locquet, M. Sciamanna, and D. S. Citrin, Optics Letters {\bf 32}, 2960 (2007).
\bibitem{Amb09} G. Ambika and R.E.Amritkar, Phys. Rev. {\bf E 79}, 056206 ( 2009).

\bibitem{Gju10} A. Gjurchinovski and V. Urumov, Phys. Rev. {\bf E 81}, 016209 (2010). 
\bibitem{Gho07} D. Ghosh, S. Banerjee and A. Roy Chowdhury, Eur.Phy. Lett. {\bf 80}, 30006 (2007).
\bibitem{Sen07} D. V. Senthilkumar and M. Lakshmanan, CHAOS {\bf 17}, 013112 (2007).

\bibitem{Zhe08} Z. Zheng, P. Yu and X. Laio, Int. J. Bif. \& Chaos {\bf 18}, 160 ( 2008).

\bibitem{Sou08}  F. O. Souza, R. M. Palhares, E. Mendes and L. Torres, Int. J. Bif. \& Chaos, {\bf 18}, 187 (2008).


\bibitem{Pyr98pre} K. Pyragas, Phys. Rev. {\bf E 58}, 3067 (1998).


\bibitem{Chen07} M. Chen and J. Kurths, Phys. Rev. {\bf E 76}, 036212 (2007).

\bibitem{Sano07} S. Sano, A. Uchida, S. Yoshimori, and R. Roy,  Phys. Rev. {\bf E 75}, 016207 (2007).
\bibitem{Tha07}  T. M. Hoang, International Journal of Electrical and Electronics Engineering {\bf 2}, 240 (2007).
\bibitem{Hoa05} T. M. Hoang, D. T. Minh, and M. Nakagawa, J. Phys. Soc. Jpn. {\bf 74}, 2374 (2005). 
\bibitem{Sha06} E.M. Shahverdiev , R.A. Nuriev , R.H. Hashimov and K.A. Shore, Chaos, Solitons \& Fractals {\bf 29}, 854 (2006).

\end{thebibliography}
\end{document}